\documentclass[aps,floatfix,superscriptaddress,twocolumn,tightenlines,amsmath,amssymb,nofootinbib,10pt,longbibliography,prl]{revtex4-1}

\usepackage[utf8]{inputenc}
\usepackage[T1]{fontenc}
\usepackage[mathscr]{euscript}
\usepackage{graphicx}
\usepackage[dvipsnames]{xcolor}
\usepackage{amsthm}
\usepackage{psfrag}
\usepackage{ifsym}
\usepackage{dsfont}
\usepackage{enumerate}
\usepackage{amsfonts}
\usepackage{multirow} 
\usepackage{amsmath}
\usepackage[sort&compress]{natbib}
\usepackage[separate-uncertainty=true]{siunitx}
\usepackage{comment}

\newcommand{\bra}[1] {\langle #1 |}
\newcommand{\ket}[1] {| #1 \rangle}

\newcommand{\ketbra}[1]{ | #1 \rangle\!\langle #1 |}
\newcommand{\Tr} {\operatorname{Tr}}
\newcommand{\expec}[1]{\left\langle #1 \right\rangle}

\newcommand{\todo}[1]{\textbf{\textcolor{blue}{ #1}}}

\begin{document}
	\title{Enhanced Multi-Qubit Phase Estimation in Noisy Environments by Local Encoding}
	
	\author{Massimiliano Proietti}
	\affiliation{
		Scottish Universities Physics Alliance (SUPA), Institute of Photonics and Quantum Sciences, School of Engineering and Physical Sciences, Heriot-Watt University, Edinburgh EH14 4AS, UK}
	
	\author{Martin Ringbauer}
	\affiliation{Institut f{\"u}r Experimentalphysik, Universit{\"a}t Innsbruck, 6020 Innsbruck, Austria}
	
	\author{Francesco Graffitti}
	\affiliation{
		Scottish Universities Physics Alliance (SUPA), Institute of Photonics and Quantum Sciences, School of Engineering and Physical Sciences, Heriot-Watt University, Edinburgh EH14 4AS, UK}
	
	\author{Peter Barrow}
	\affiliation{
		Scottish Universities Physics Alliance (SUPA), Institute of Photonics and Quantum Sciences, School of Engineering and Physical Sciences, Heriot-Watt University, Edinburgh EH14 4AS, UK}
	
	\author{Alexander Pickston}
	\affiliation{
		Scottish Universities Physics Alliance (SUPA), Institute of Photonics and Quantum Sciences, School of Engineering and Physical Sciences, Heriot-Watt University, Edinburgh EH14 4AS, UK}
	
	\author{Dmytro Kundys}
	\affiliation{
		Scottish Universities Physics Alliance (SUPA), Institute of Photonics and Quantum Sciences, School of Engineering and Physical Sciences, Heriot-Watt University, Edinburgh EH14 4AS, UK}
	
	\author{Daniel Cavalcanti}
	\affiliation{ICFO-Institut de Ciencies Fotoniques, The Barcelona Institute of Science and Technology,  08860 Castelldefels (Barcelona), Spain}
	
	\author{Leandro Aolita}
	\affiliation{Instituto de F\'isica, Universidade Federal do Rio de Janeiro, P. O. Box 68528, Rio de Janeiro, RJ 21941-972, Brazil}
	
	\author{Rafael Chaves}
	\affiliation{International Institute of Physics, Federal University of Rio Grande do Norte, 59070-405 Natal, Brazil}
	\affiliation{School of Science and Technology, Federal University of Rio Grande do Norte, 59078-970 Natal, Brazil}
	
	\author{Alessandro Fedrizzi}
	\affiliation{
		Scottish Universities Physics Alliance (SUPA), Institute of Photonics and Quantum Sciences, School of Engineering and Physical Sciences, Heriot-Watt University, Edinburgh EH14 4AS, UK}

	\begin{abstract}
		The first generation of multi-qubit quantum technologies will consist of noisy, intermediate-scale devices for which active error correction remains out of reach. To  exploit such devices, it is thus imperative to use passive error protection that meets a careful trade-off between noise protection and resource overhead.
		Here, we experimentally demonstrate that single-qubit encoding can significantly enhance the robustness of entanglement and coherence of four-qubit graph states against local noise with a preferred direction. In particular, we explicitly show that local encoding provides a significant practical advantage for phase estimation in noisy environments. This demonstrates the efficacy of local unitary encoding under realistic conditions, with potential applications in multi-qubit quantum technologies for metrology, multi-partite secrecy and error correction.
	\end{abstract}
	
	\maketitle
	
	\emph{Introduction---}Quantum systems are notoriously fragile due to unavoidable interactions with their environment~\cite{Zurek03}, resulting in decoherence that grows exponentially with system size. This represents a major roadblock for quantum computing~\cite{Nielsen}, quantum communication~\cite{Aschauer2002} and quantum metrology~\cite{Giovannetti2011}, rendering noise mitigation~\cite{Freedman2003,Lidar1998,Kim2012} indispensable. Quantum error correction (QEC)~\cite{Shor1995,Calderbank1996,Steane1996} schemes in principle achieve full protection against decoherence. However, daunting experimental requirements on the single qubit noise rate and large resource overheads~\cite{Aharonov2008} make QEC a long-term vision.
	
	A complementary approach, expected to play a central role in near-term quantum technologies~\cite{Preskill18}, is to relax the fault-tolerance requirement against arbitrary noise aiming instead at enhanced robustness of quantum systems, under experimentally relevant conditions. One of the dominant types of noise is local dephasing along a privileged direction~\cite{Noda1986,Leibfried2003,Schindler2013}. There, simple single-qubit unitary encoding can drastically improve the resilience of quantum resources~\cite{Chaves2012} such as multi-qubit Greenberger-Horne-Zeilinger (GHZ)~\cite{Greenberger1990} states, where an exponential decay of entanglement~\cite{Aolita2008,Aolita2015} can be turned into a linear decay~\cite{Chaves2012}. This improvement is crucial for metrology applications such as phase estimation in noisy environments~\cite{shaji2007qubit,Chaves2013}, whereby the otherwise optimal phase sensitivity of GHZ states~\cite{giovannetti2004quantum} become asymptotically bounded by a constant~\cite{demkowicz2012elusive}.

	Here, in a state-of-the-art 4-photon experiment at telecom wavelength, we report an in-depth study of the enhanced noise resilience that can be gained from local encoding~\cite{Chaves2012}. Using symmetric informationally complete (SIC)~\cite{Renes2004SIC} tomographic techniques, we quantify the noise resilience of quantum resources such as coherence and entanglement for all local-unitarily inequivalent classes of 4-qubit graph states~\cite{Hein2005,hein2004multiparty}. 
	Finally, as quantified by the experimental phase variance and the quantum Fisher information~\cite{Holevo2011}, we observe that our encoding provides a clear improvement of the 4-qubit GHZ states usefulness for noisy quantum phase estimation. Notably, even under full dephasing, where the entanglement is always zero, the states nonetheless remain useful, hinting at a key role played by coherence, shown instead to be independent from the noise when the encoding is applied.
	
	
	\emph{Multi-Qubit Robustness by Local Encoding---}
	Consider an ideal quantum system subjected to local dephasing noise, before being used for an information-processing task such as phase estimation. Such is the case, for example, when the system crosses a noisy region before the protocol happens or if its implementation is much faster than the dephasing timescale. We note that the results are the same when the noise acts during the phase estimation task, here however, we keep them separate for simplicity. We aim to encode the system before the noise acts in order to increase its resilience against dephasing with as simple an encoding as possible, and then decode the system before it is being used. Taking the dephasing to act on the state $\rho$ in the computational basis $\{\ket{0},\ket{1}\}$, the single-qubit dephasing channel is given by
	\begin{equation}
		\mathcal{D}(\rho) \doteq \left(1-\frac{p}{2} \right)
		\rho+\frac{p}{2}\sigma_{z} \rho \sigma_{z},
		\label{eq:dephasing}
	\end{equation}
	where $\sigma_{z}=\ketbra{0}-\ketbra{1}$ is the $Z$ Pauli matrix, and $0\leq p\leq 1$ quantifies the noise strength from no noise ($p=0$) to full dephasing ($p=1$). Consider now, for instance, an $N$-qubit GHZ state defined as
	\begin{equation}
		\ket{GHZ_{N}} \doteq\frac{1}{\sqrt{2}}\left(\ket{0} ^{\otimes N} +\ket{1} ^{\otimes N}\right)
		\label{eq:ghzstate}
	\end{equation}
	The entanglement of GHZ states under single-qubit dephasing is known to decay exponentially with $N$. More precisely, for the dephased GHZ state $\rho_N (p)\doteq\mathcal{D}^{\otimes N}(\ketbra{GHZ_{N}})$ it holds that $E\left(\rho_N (p)\right)\leq (1-p)^N E\left(\rho_N (0)\right)$ for any convex entanglement quantifier $E$~\cite{Aolita2008,Aolita2015}.
	However, this scaling can be drastically improved~\cite{Chaves2012} by encoding the state using local Hadamard gates $H$, defined by $H\ket{0}\doteq\ket{+}$ and $H\ket{1}\doteq\ket{-}$ with $\ket{\pm}\doteq\frac{1}{\sqrt{2}}(\ket{0}\pm\ket{1})$. The resulting encoded GHZ state:
	\begin{equation}
		\ket{{{GHZ}_{N}^{enc}}}\doteq H^{\otimes N} \ket{{{GHZ}_{N}}}=\frac{1}{\sqrt{2}}\left(\ket{+}^{\otimes N} +\ket{-}^{\otimes N}\right)
		\label{eq:ghzstatetrans}
	\end{equation}
	has the same entanglement properties, yet its entanglement decay rate becomes independent of $N$ and linear in $p$. Formally, the dephased encoded state $\rho^{enc}_N (p)\doteq\mathcal{D}^{\otimes N}(\ketbra{GHZ_{N}^{enc}})$ satisfies the bound $E\left(\rho^{enc}_N (p)\right) \geq  E\left(\rho_2(p)\right)$ for all $N$ and thus possesses at least as much resilience as the two-qubit state $\ket{GHZ_2}$. A similar enhancement can be extended for arbitrary graph states~\cite{hein2004multiparty}, which play a crucial role in measurement-based quantum computing and quantum error correction, see Supplementary Materials (SM) for details. 
	\begin{figure}[t!]
		\begin{center}
			\includegraphics[width=0.95\columnwidth]{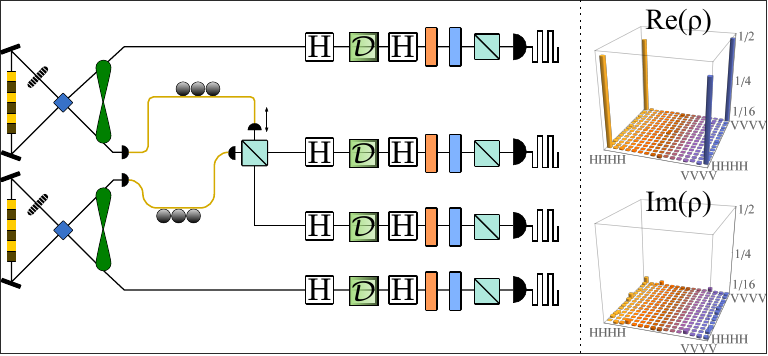}
		\end{center}
		\vspace{-2em}
		\caption{\textbf{Experimental Setup}. Preparation of 4-qubit GHZ states and encoding stage. Pairs of photons at $1550$~nm are generated via spontaneous parametric downconversion in periodically-poled KTP (PPKTP) crystals. The GHZ state is obtained from interfering photons from two entangled pairs on a polarizing beam splitter (PBS). The state is then locally encoded, dephased, decoded, and measured using a combination of quarter-waveplate (QWP), half-waveplate (HWP), PBS, and superconducting nanowire single photon detectors (SNSPDs) with four-fold coincidence detection. On the right, the real and imaginary part of the experimental density matrices (without dephasing) are shown.}
		\label{fig:Figure1}
	\end{figure}

	\emph{Experimental setup---}We now test these passive error protection techniques in a state-of-the-art photonic platform, Fig.~\ref{fig:Figure1}. Qubits are encoded in the horizontal $\ket{h}=\ket{0}$ and vertical $\ket{v}=\ket{1}$ polarization states of single photons. These are generated at $\SI{1550}{nm}$ via collinear type-II spontaneous parametric down-conversion in a \SI{22}{mm} long periodically-poled KTP (PPKTP) crystal, pumped with a \SI{1.6}{ps} pulsed laser at \SI{775}{nm}~\cite{graffitti2018}. After spectral filtering with a bandwidth of \SI{3}{nm}, the source generates $\sim 3075$ pairs/mW/s with a symmetric heralding efficiency of $\sim 55\%$. Embedding the crystal within a Sagnac interferometer~\cite{Fedrizzi2007Sagnac} enables the generation of high-quality entangled states of the form
	\begin{equation}
		\ket{\psi^{-}}=\frac{1}{\sqrt{2}}\left(\ket{h}\ket{v}-\ket{v}\ket{h}\right) ,
		\label{psim}
	\end{equation}
	with typical fidelities $F(\rho_e,\rho_t)= (Tr[\sqrt{\sqrt{\rho_t}\rho_e\sqrt{\rho_t}}])^{2}=99.62^{+0.01}_{-0.04}\%$ where $\rho_e$ and $\rho_t$ are the experimental and target state respectively. The measured purity is $P=99.34^{+0.01}_{-0.09}\%$ and entanglement as measured by the concurrence~\cite{Hill1997} is $\mathcal{C}=99.38^{+0.02}_{-0.10}\%$. 
	The photons are detected using superconducting nano-wire single-photon detectors (SNSPDs) with an efficiency of $\sim 80\%$ and processed using a time-tagging module with a resolution of \SI{156}{ps}. Using two such photon-pair sources in the setup of Fig.~\ref{fig:Figure1}, we can prepare the 4-qubit GHZ state $\ket{GHZ_4}$ of Eq.~\eqref{eq:ghzstate} by subjecting one photon of each entangled pair to nonclassical interference on a polarizing beam splitter (PBS), which transmits horizontal and reflects vertically polarized photons. This implements a so-called type-I fusion gate~\cite{Browne2005} for which we achieved a visibility of $91.80^{+1.73}_{-1.73} \%$, translating into a purity of $P=87.09^{+1.15}_{-2.18\%}$ and fidelity of $F=92.53^{+0.63}_{-1.23}\%$ for the 4-qubit GHZ state. The states are generated at a measured rate of \SI{47.6}{Hz} using \SI{60}{mW} pump power.

	Single qubit dephasing of Eq.~\eqref{eq:dephasing} is experimentally implemented in a controllable manner by applying individual phase flips, using HWPs, to a fraction of the runs weighted by the dephasing strength $p$. The density matrices of the experimentally generated states are then reconstructed using maximum-likelihood quantum state tomography. The tomography was performed using the set of symmetric informationally complete (SIC) measurements~\cite{Renes2004SIC}, which reduces the number of measurements compared to the standard Pauli basis by a factor $(2/3)^N$, leading to improved precision at equal acquisition time.
	
	\begin{figure}[t!]
		\begin{center}
			\includegraphics[width=0.8\columnwidth]{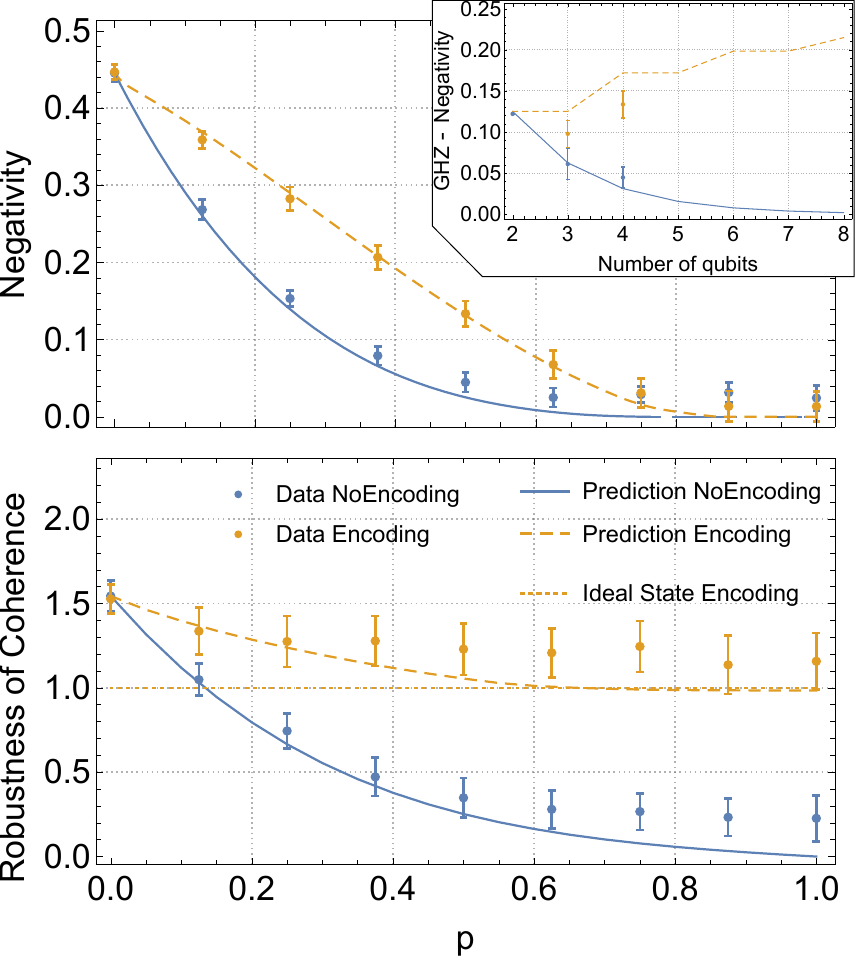}
		\end{center}
		\vspace{-2em}
		\caption{\textbf{Resilience enhancement of negativity in the partition $(1|234)$ and the robustness of coherence $R_{C_1}$.} Shown is the negativity (top) and coherence (bottom) of the GHZ with (dashed-orange) and without (solid-blue) encoding. The solid (dashed) lines depict the theoretical predictions with the experimental bare (encoded) input state. In the top-right inset the trend of the GHZ negativty is shown in terms of the number of qubits and at fixed noise $p=0.5$. With the encoding proposed, the entanglement is best protected when the number of qubits increases. For the coherence only, the theory prediction starting with an ideal input encoded state is shown with a dotted-orange curve. Note that experimental imperfections tend to lead to additional coherence terms compared to the ideal GHZ state. The robustness of coherence reflects this as higher initial values of coherence and non-vanishing coherence for all dephasing strengths. Error bars represent $3\sigma$ statistical confidence regions obtained from a Monte-Carlo routine taking into account the Poissonian counting statistics.}
		\label{fig:NegativityResults}
	\end{figure}

	\begin{figure*}[t!]
		\begin{center}
			\includegraphics[width=0.9\textwidth]{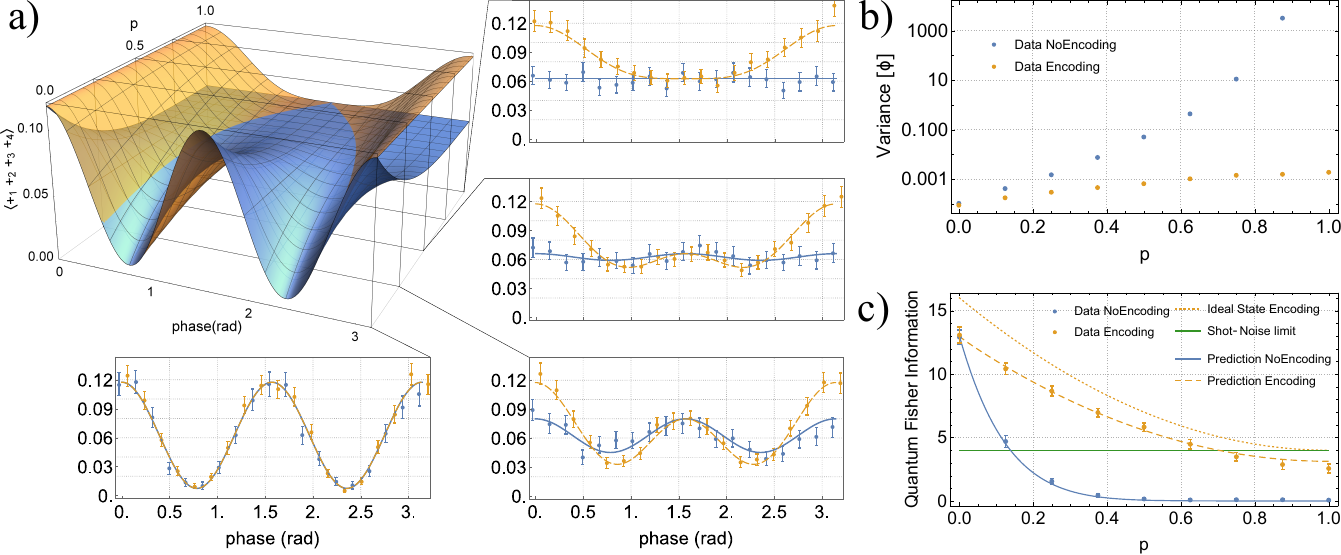}
		\end{center}
		\vspace{-1em}
		\caption{\textbf{Phase estimation with and without encoding.} \textbf{a)} Expectation value $\expec{+_1+_2+_3+_4}$ as a function of phase and amount of noise $p$, for a locally encoded (orange) and a non-encoded (blue) 4-qubit GHZ state. In particular, for values of $p=0,0.25,0.5,1$ the experimentally measured expectation values as a function of the phase are shown. The theoretical predictions are shown as blue-solid (no encoding) and orange-dashed (encoding) curves, and error bars indicate $3\sigma$ statistical uncertainty regions obtained from a Monte Carlo resampling of our Poisson counting statistics. In the absence of noise ($p=0$) there is no difference between encoded and bare states. With increasing dephasing, however, the advantage of the encoding becomes clear in that the expectation values for bare states decay to zero, but remain non-zero for all $p$ if the local encoding is used. \textbf{b)} Robustness enhancement of the quantum Fisher information. QFI of the encoded (dashed-orange) and non-encoded (solid-blue) states, compared with the shot-noise limit (solid green). Without encoding, the GHZ state loses its advantage already in the low-noise regime. In contrast, as shown in the figure, the encoding preserves the QFI for all values of dephasing, ideally (dotted-orange), and up to $p=0.6$ experimentally. \textbf{c)} Comparison of the phase variance without (blue) and with (orange) encoding, for different noise strengths. Notably, with encoding, the variance observed is up to 2 orders of magnitude smaller than the case without, where the error on the inferred phase diverges with increasing noise.}
		\label{fig:PhaseEstimationResults}
	\end{figure*}

	\emph{Experimental noise protection---}In this section we present the experimental results attesting the capability of the method proposed, whose application in phase estimation is shown in the next section.
	We investigate the effect of the encoding proposed on two paradigmatic quantum resources: quantum entanglement and coherence. The former is quantified by the negativity~\cite{Vidal2002}, the latter instead using the recently developed resource theory of multilevel coherence~\cite{Napoli2016,Ringbauer2017}, see SM for details. The results for both these figures of merit are shown in Fig.~\ref{fig:NegativityResults}. The top panel shows that the negativity of the encoded $\ket{GHZ_4^{enc}}$ states is significantly more resilient against dephasing than the bare states. Moreover, the inset in Fig.~\ref{fig:NegativityResults} shows how the enhancement for a fixed amount of dephasing becomes more significant as the number of qubits increases, instead of the exponential decay observed for the bare states~\cite{Chaves2012}. As shown in the SM, such noise resilience is notably achieved by increasing the amount of entanglement with the environment. In fact, the encoded states experience a higher loss of purity than the non-encoded ones. 
	
	The bottom panel of Fig.~\ref{fig:NegativityResults}, shows how coherence of the encoded states not only is protected from the action of noise but distinctly manifests independence from it. Non-encoded states, on the other hand, show an exponential decay. Intuitively, this may be understood based on the distribution of coherence within the state. Concentrating all coherence on two terms (coherence rank 2), such as the bare GHZ$_4$ state leaves the state vulnerable to dephasing. In contrast, maximally spreading it out (coherence rank $2^N$), as in the encoded state, achieves increased resilience. In the latter case indeed, the decoding map (a non-free operation in the resource theory of coherence) can under certain conditions recover a significant amount of coherence, see SM for details. We remark that, multi-level coherence is independent of entanglement measures and can unlock information on the encoding effects otherwise inaccessible. This, as we will see in the next section, provides useful insights on phase estimation in noisy environment. Finally, noise protection, is also achieved for the linear cluster state, see SM.

	\emph{Enhanced phase estimation---}We now exploit our passive error correction for quantum metrology, by performing a 4-qubit phase estimation task~\cite{MetrologyReview} in a noisy environment. The goal is to estimate an unknown phase $\phi$ imparted on a probe state $\rho$ by the unitary $U_{\phi}=e^{-\frac{i}{2}\phi \sigma_z}$ by measuring the evolved state $\rho_{\phi}\doteq U_{\phi}^{\otimes N}\rho {U_{\phi}^{\dagger}}^{\otimes N}$. It is well known that GHZ states are optimal for phase estimation~\cite{Giovannetti2011}. In the presence of dephasing noise, however, this task becomes much more challenging, and different inequivalent strategies can be devised~\cite{demkowicz2014using,Pirandola2017}. We now show how our local encoding can significantly enhance the metrology performance of 4-qubit GHZ state under such conditions. To assess the performance of phase estimation we study the expectation value: 
	\begin{equation*}
		\Tr[\rho_{\phi}\left(\ket{+}\bra{+}\right)^{\otimes 4}]_{\textsc{ghz}} =\frac{1}{16} \left((p-1)^4 \cos (4 \phi )+1\right) ,
	\end{equation*}
	for noise of strength $p$, showing that for maximal dephasing i.e. $p=1$, no phase information can be recovered. Conversely, if the encoding of Eq.~\eqref{eq:ghzstatetrans} is used, we find improved performance for all $p$. Most strikingly, the encoding preserves phase sensitivity even under full dephasing:
	\begin{equation*}
		\Tr[\rho_{\phi}\left(\ket{+}\bra{+}\right)^{\otimes 4}]_{\textsc{ghz}^{enc}} =\frac{1}{128} (4 \cos (2 \phi )+\cos (4 \phi )+11) .
	\end{equation*}
	Although entanglement is recognized as a fundamental resource for phase estimation~\cite{TothEntanglementMetrology,SmerziEntMetrology}, is remarkable to note that phase sensitivity is observed even in the full dephasing regime, whereby the entanglement is always zero, even for the encoded states. It follows that the phase sensitivity observed is instead provided by the coherence only, which as we have previously seen is left untouched by the dephasing. Only when both coherence and entanglement are zero (as for the non-encoded states) the phase sensitivity is completely suppressed. This suggests, at least for this instance, the coherence to be a useful resource whereby the entanglement is not.

	Experimentally, we applied a phase-shift $\phi\in[0,\pi]$ to each qubit, by shifting the measurement's waveplates accordingly and reconstructed the expectation values $\Tr[\rho_{\phi}\left(\ket{+}\bra{+}\right)^{\otimes 4}]$ as a function of $\phi$ for a range of $p$, see Fig.~\ref{fig:PhaseEstimationResults}a. The results clearly show a steeper slope of $\Tr[\rho_{\phi}\left(\ket{+}\bra{+}\right)^{\otimes 4}]$  for the encoded state compared to the bare state for all non-zero values of $\phi$. This directly translates into a more sensitive phase estimator in the encoded case. Moreover, we emphasize that the encoded fringes preserve at least half the visibility of the $p=0$ case, even for $p=1$ where instead, without our encoding, the bare fringes flatten to a constant value. In other words, whereby phase estimation would be normally impossible, our encoding makes it feasible again.
	
	This qualitative behavior is turned into a quantitative result by measuring the experimental variance of the estimated phase in the point where the fringes are the steepest for different values of $p$, according to
	\begin{equation}
		Var[\phi]=\frac{Var[\epsilon]}{|\frac{d}{d\phi}\epsilon|^{2}},
	\end{equation}
	where $\epsilon$ is the measured average value of our estimator $\epsilon\equiv\expec{+_{1}+_{2}+_{3}+_{4}}$. The results are shown in Fig.~\ref{fig:PhaseEstimationResults}b.
	
	More in general however, the primary figure of merit in quantum metrology is the so called \emph{quantum Fisher information} (QFI)~\cite{Holevo2011}. Indeed, in the noiseless case, the statistical deviation $\delta\phi$ in the estimation of $\phi$, is bounded as $\delta\phi\geq1/\left(\sqrt{\nu\mathcal{F}(\rho_{\phi})}\right)$~\cite{Braunstein1994}, where $\nu$ is the number of repetition runs in the estimation and $\mathcal{F}(\rho_{\phi})$ is the QFI, measuring the maximum amount of information about $\phi$ that can be extracted from $\rho_{\phi}$. For separable states $\mathcal{F}(\rho_{\phi})\leq N$, namely the QFI is bounded by the shot-noise limit (SNL), while for GHZ states the QFI attains the optimal value $\mathcal{F}_{\text{max}}=N^{2}$, known as the Heisenberg limit. On the other hand, the QFI of a locally dephased GHZ state $\mathcal{F}(\rho_N(p))=N^2(1-p)^{2N}$, indicates that for fixed noise strength $p$ the precision of the estimate of $\phi$ decreases exponentially with $N$. Such drastic decay is turned into a quadratic one for the encoded GHZ state, $\mathcal{F}(\rho_N^T (p))=N^{2}(1-p)^{2}+4N\big(1-\frac{p}{2}\big)\frac{p}{2}$. The Fisher information is measured experimentally for both the encoded and non-encoded density matrices~\cite{Liu2014}, see Fig.~\ref{fig:PhaseEstimationResults}c. In absence of noise we experimentally observe a value close to the Heisenberg limit $N^{2}=16$ exponentially turned to $0$ when the encoding is not applied. On the other hand, encoded GHZ states preserve their quantum advantage for significantly high noise strengths.

	\emph{Conclusions---}
	We have shown that, given knowledge of the dominant noise sources in the experiment, protection of quantum resources without complex encoding and additional overhead in the number of physical qubits, is feasible in practice and can enable significant performance improvements. We revealed different behaviors of graph states for all the quantum figure of merits under study, highlighting the importance of a deep understanding of state dynamics in the multi-qubit scenario. In particular, we exploited our method in one of the most common quantum information tasks i.e. phase estimation. Adding dephasing noise, we simulated a realistic implementation of such protocol where without any action no quantum advantage would be observed. With our encoding, we instead observe phase-sensitivity up to the full-noise regime where only coherence and not entanglement might be the useful resource for such task. In conclusion, we successfully proposed an alternative route for noise-protection which might be further exploited as far as active multi-qubit quantum error correction remains out of reach for near-term technology.

	\begin{acknowledgments}
		This work was supported by the UK Engineering and Physical Sciences Research Council (grant number EP/N002962/1), the Ram\'on y Cajal fellowship (Spain), Spanish MINECO (QIBEQI FIS2016-80773-P and Severo Ochoa SEV-2015-0522), the AXA Chair in Quantum Information Science, Generalitat de Catalunya (SGR875 and CERCA Programme), Fundaci\'{o} Privada Cellex and ERC CoG QITBOX. FG acknowledges studentship funding from EPSRC under grant no. EP/L015110/1. MR acknowledges funding from the European Union's Horizon 2020 research and innovation programme under the Marie Sk\l{}odowska-Curie grant agreement No 801110 and the Austrian Federal Ministry of Education, Science and Research (BMBWF).
		LA acknowledges financial support from the Brazilian agencies CNPq (PQ grant No. 311416/2015-2 and INCT-IQ), FAPERJ (JCN  E-26/202.701/2018), CAPES (PROCAD2013),  FAPESP, and the Serrapilheira Institute (grant number Serra-1709-17173). RC acknowledges the Brazilian ministries MEC and MCTIC, funding agency CNPq (PQ grants No. 307172/2017-1 and No 406574/2018-9 and INCT-IQ), the John Templeton Foundation via the grant Q-CAUSAl No. 61084 and the Serrapilheira Institute (grant number Serra-1708-15763).
	\end{acknowledgments}

	
	%

	\newpage
	\clearpage
	\renewcommand{\theequation}{S\arabic{equation}}
	\renewcommand{\thefigure}{S\arabic{figure}}
	\renewcommand{\thetable}{\Roman{table}}
	\renewcommand{\thesection}{S\Roman{section}}
	\setcounter{equation}{0}
	\setcounter{figure}{0}
	\section{Supplementary Material}
	
\begin{figure}[t!]
		\begin{center}
			\includegraphics[width=0.9\columnwidth]{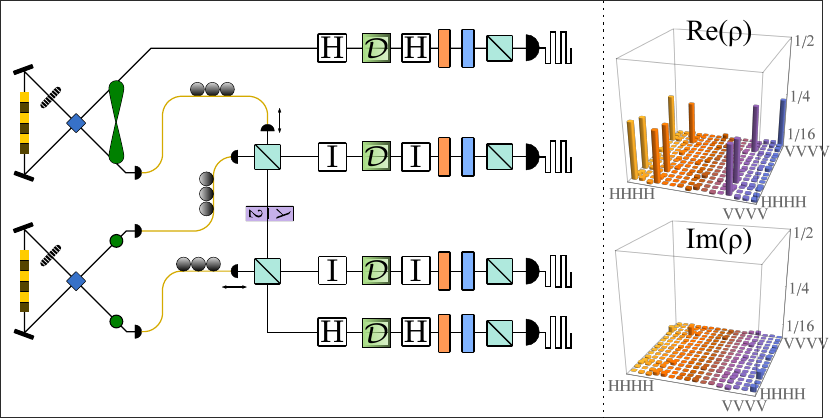}
		\end{center}
		\caption{\textbf{Linear Cluster experimental setup}. Pairs of photons at $1550$~nm are generated via spontaneous parametric downconversion in periodically-poled KTP (PPKTP) crystals. The encoded linear cluster is obtained from two successive interference of one photon from an entangled pair with two diagonally polarized single photons. The state is then locally encoded, dephased, decoded, and measured using a combination of quarter-waveplate, half-waveplate, PBS, and superconducting nanowire single photon detectors with four-fold coincidence detection. On the right, the real and imaginary part of the experimental density matrices (without dephasing) are shown.}
		\label{fig:FigureS1}
	\end{figure}

\begin{figure}[t!]
		\begin{center}
			\includegraphics[width=0.85\columnwidth]{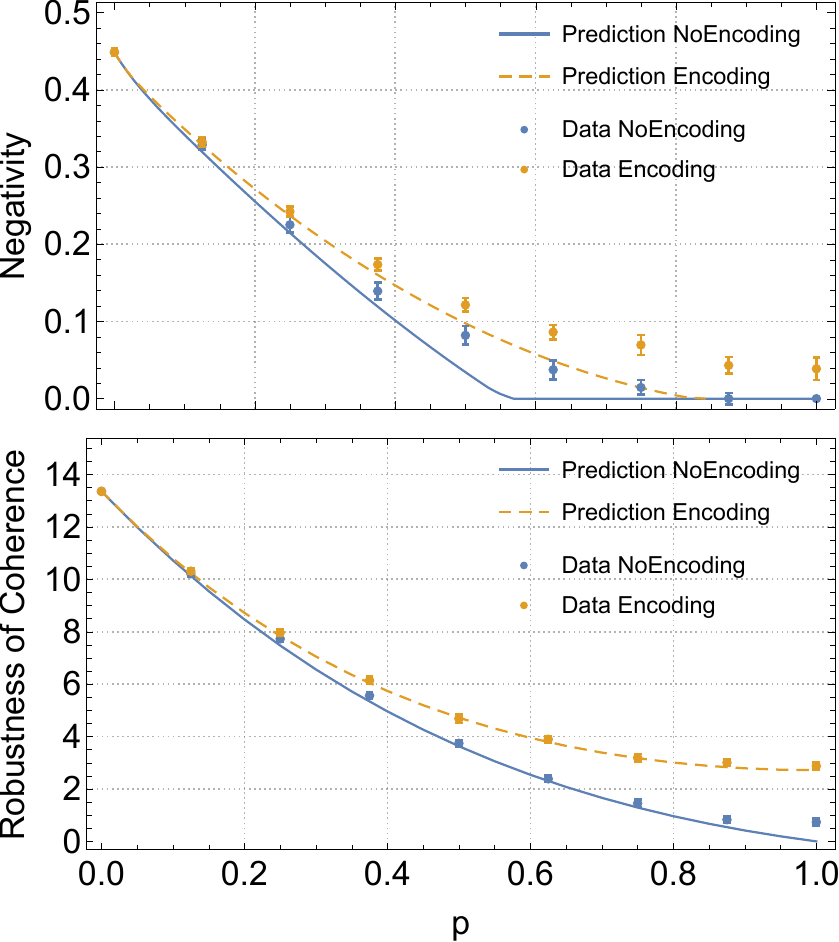}
		\end{center}
		\caption{\textbf{Resilience enhancement of negativity in the partition $(1|234)$ and the robustness of coherence $R_{C_1}$.} Shown is the negativity (top) and coherence (bottom) of the CL4 with (dashed-orange) and without (solid-blue) encoding. The solid (dashed) lines depict the theoretical predictions with the experimental bare (encoded) input state. Notably, for high values of noise, a zero-negativity observed without encoding is turned into a non-zero value . Error bars represent $3\sigma$ statistical confidence regions obtained from a Monte-Carlo routine taking into account the Poissonian counting statistics.}
		\label{fig:FigureS2}
	\end{figure}

	\emph{Extension to linear cluster graph state---}Graph states are a sub-class of multi-qubit states which can be expressed by means of a graph, where vertices represent qubits and links entangling interactions. These play a fundamental role from quantum computation to quantum error correction. For $N=4$ qubits there exist only two classes of connected graph states in-equivalent under local unitary transformation~\cite{hein2004multiparty} i.e. a state in one class can not be mapped to a state in the other by means of any local unitary operation. The paradigmatic representatives of these two families are the GHZ state $\ket{GHZ_4}$ and the linear cluster state $\ket{CL_{4}}$, with the latter commonly defined as
	\begin{equation}
		\begin{split}
			\label{eq:noEnclinearCluster}
			\ket{CL_{4}}\doteq &CZ_{1,2}CZ_{2,3}CZ_{3,4}HHHH\ket{0000}=\\
			= &\frac{1}{2}(\ket{+00+}+\ket{-10+}+\\
			&\hspace{2.5cm}+\ket{+01-}-\ket{-11-}),
		\end{split}
	\end{equation}
	where the tensor product is omitted and $CZ_{i,j}$ is a controlled-Z gate~\cite{Nielsen} on qubits $i$ and $j$. This expression corresponds to the graph state represented as a linear chain with qubits $1,4$ as external vertices. By investigating with our protocol both the GHZ and the linear cluster quantum states, we could in essence covers all 4-qubit graph states. The optimal local-unitary encoding for $\ket{CL_{4}}$ turns out to be $H\otimes \openone\otimes \openone\otimes H$, with $\openone$ the single-qubit identity operator, resulting in the state
	\begin{equation}
		\label{eq:enclinearCluster}
		\ket{CL_{4}^{enc}}=\frac{1}{2}\left(\ket{0000}+\ket{1100}+\ket{0011}-\ket{1111}\right)
	\end{equation}
	This state can be generated experimentally by subjecting one photon of an entangled pair to two sequential fusion gates with uncorrelated single photons in the state $\ket{+}$ and with a Hadamard gate in between, see Fig.~\ref{fig:FigureS1}. The purity and fidelity of the obtained state are $\mathcal{P}=81.77^{+0.65}_{-0.85\%}$ and $\mathcal{F}=89.03^{+0.38}_{-0.60}\%$ respectively. As for the GHZ case we compute the negativity in the partition $(1|234)$ and the robustness of multilevel coherence. As shown in Fig.~\ref{fig:FigureS2} the encoding provided enhances the resilience of linear cluster states against the action of dephasing. Moreover, it is even possible to qualitatively change the behavior from finite-time disentanglement to infinite-time disentanglement. Regarding the coherence, contrary to the GHZ case, a constant behavior cannot be achieved, due to subtle differences in the structure of the states. Nonetheless protection is observed for all the values of dephasing. We note that, since contrary to the GHZ state the linear cluster state is known to be non-optimal for phase estimation, we limited our study to the case of entanglement and coherence enhancement of locally encoded linear cluster states.
	
	Finally, for both the GHZ state and Linear Cluster, we consider all the 1-vs-rest partitions $(i|jkl)$ and 2-vs-2 bipartitions : $(12|34), (13|24), (14|23)$. The encoded 4-GHZ state is optimally protected in both all the 1-vs-rest and 2-vs-2 partitions. The scenario is more complex when instead we study the linear cluster, in fact there always exists at least one 1-vs-rest partition where the protection is optimal. However, the partition depends from the encoding chosen. Nevertheless, encoding providing protection for all the 1-vs-rest partitions exist, but the protection is not optimal.
	
\begin{figure}[t!]
		\begin{center}
			\includegraphics[width=0.9\columnwidth]{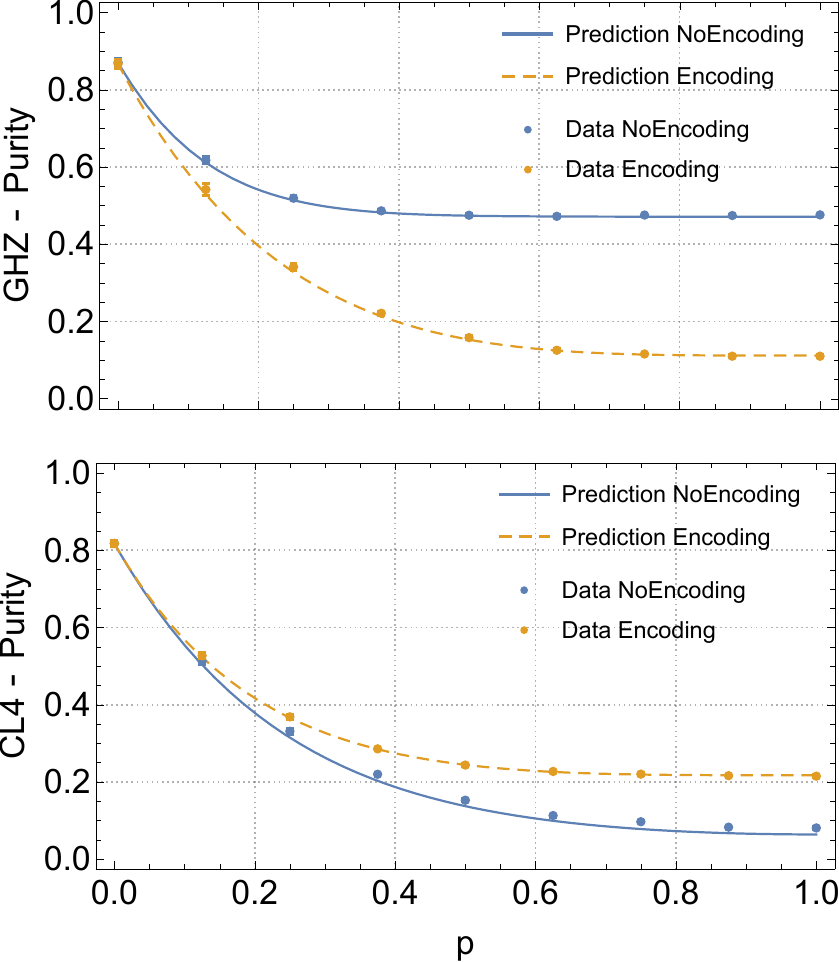}
		\end{center}
		\caption{\textbf{Experimental results for the purity under local encoding}. Shown is the purity of the GHZ (top) and linear cluster (bottom) states with (dashed-orange) and without (solid-blue) encoding. The solid (dashed) lines depict the theoretical predictions with the experimental bare (encoded) input state. Interestingly, conversely to the GHZ case, the purity of the linear cluster is enhanced by the local encoding. The $3\sigma$ error bars, are smaller than the symbol size.}
		\label{fig:FigureS3}
	\end{figure}

	\emph{Evolution of the purity and entanglement entropy with the environment---}Intuitively, one would assume that the noise resilience is achieved by reducing the amount of entanglement with the environment that is generated (thus increasing the state's purity). Surprisingly, however, the results of Fig.~\ref{fig:FigureS3} show that, for the GHZ case, the opposite is true and the encoded states experience a higher loss of purity than the non-encoded ones.
	
	Purity of a density matrix $\rho$ is given by $Tr[\rho^{2}]$ and can be expressed in terms of the eigenvalues $\lambda_{k}$ as $\mathcal{P}=\sum_{k}\lambda^{2}_{k}$. The evolved dephased bare GHZ state has eigenvalues 
	\begin{align}
		&\lambda_0=(1/2)(1-(1-p)^{N})\\
		&\lambda_1=1-\lambda_0
	\end{align}
	leading to $P(\rho^N(p))=\frac{1}{2}\left( 1+ (1-p)^{2N}\right)$. On the other hand, the encoded state has eigenvalues given by
	\begin{equation}
		\lambda_k=(1-p/2)^{N-k}(p/2)^{k}+(1-p/2)^{k}(p/2)^{N-k}
	\end{equation}
	with $0 \leq k\leq N-1$  (each with a degeneracy of $\binom{N-1}{k}$) leading to a purity $P(\rho^N_T (p))=\left( p^N(1-\frac{p}{2})^N+ (1-p+\frac{p^2}{2})^{N}\right)$. Although both purities decay exponentially with $N$, the purity of the bare state tends to $1/2$, while the purity of encoded state tends to $2^{1-N}$. The entanglement between system and environment can also be quantified via the von Neumann entropy of the system
	\begin{equation}
		S(\rho)=\mathrm{Tr}(\rho \log_2 \rho)=-\sum_{k} \lambda_k \log_2 \lambda_k.
	\end{equation}
	For this quantity, closed formula expressions are not any longer possible  but one can easily see some interesting properties. For the bare GHZ state the entanglement entropy tends to $S(\rho_N (p))=1$ while for the encoded state $S(\rho_N^T (p))=N-1$ as $p \rightarrow 1$. Moreover, for any $p>0$ it follows that $S(\rho_N^T (p)) > S(\rho_N (p))$, that is, at all times of the noisy evolution the encoded and more robust state is surprisingly more entangled with the environment.
	
	Intuitively, this can be understood as a consequence of the special structure of the GHZ state, which even after full dephasing retains classical correlations that manifest in relatively high residual purity. In the encoded case, the coherence is more distributed, thereby reducing the resilience of the state's purity. In the case of the linear cluster, on the other hand, the bare state features uniformly distributed populations, while the encoded state is sparser. As a consequence, the optimal encoding protects both entanglement \emph{and} purity.
	
\begin{figure}[t!]
		\begin{center}
			\includegraphics[width=0.9\columnwidth]{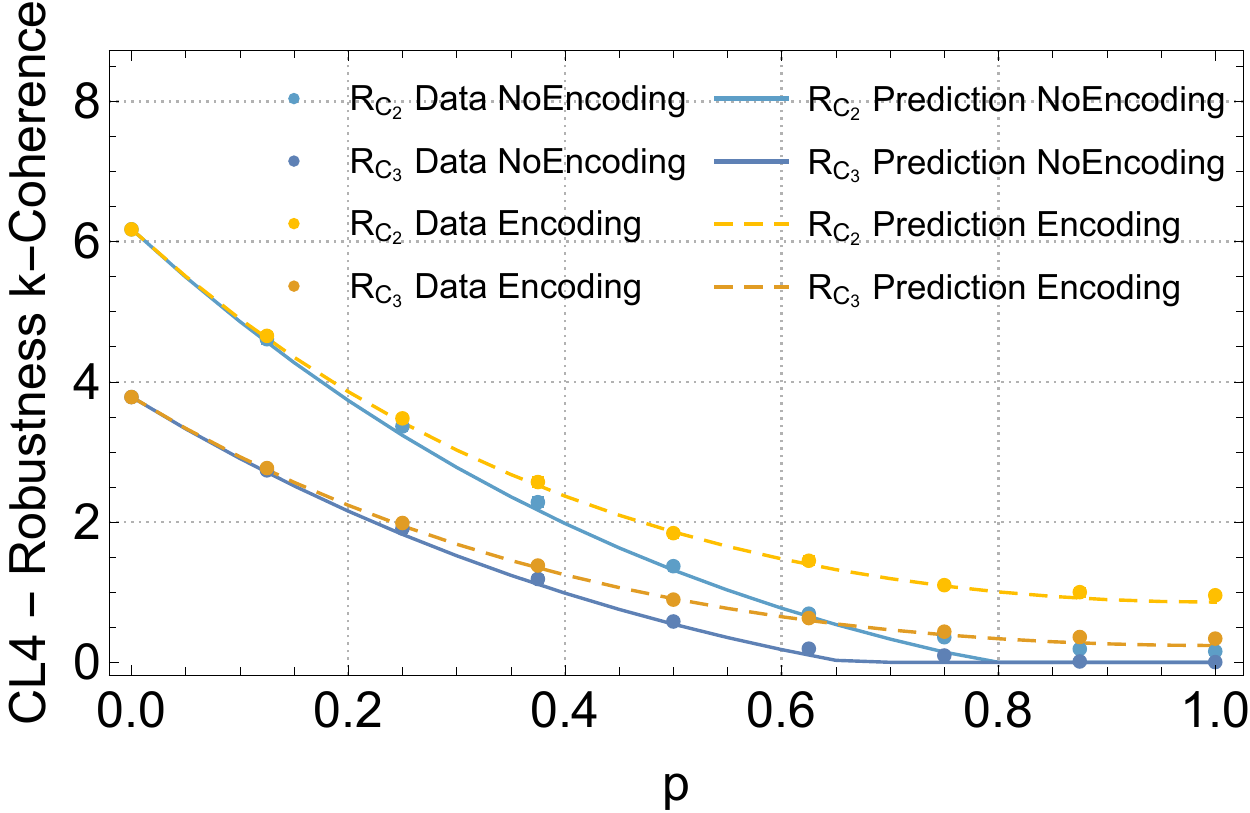}
		\end{center}
		\caption{\textbf{Robustness of multilevel coherence}. Shown is the robustness of 3-level ($R_{C_2}$) and 4-level ($R_{C_3}$) coherence the bare (blue) and encoded (orange) linear cluster. Similar to the results in the main text, the encoding provides enhanced protection of coherence at all levels and for all values of $p$. Experimental data is shown as dots with $3\sigma$ error bars.}
		\label{fig:FigureS4}
	\end{figure}

	\emph{Robustness of multilevel coherence---}Following Ref.~\cite{Ringbauer2017}, we now review the concept of multilevel quantum coherence as a fine-grained quantifier for the amount of coherence present in a given quantum state. We consider composite $N$-qubit systems and measure coherence with respect to the computational basis $\{\ket{0},\ket{1}\}^{\otimes N}$. In order to capture the structure of coherence in such a system, one first defines the following sets of states
	\begin{equation}
		\mathcal{C}_k := \operatorname{conv}\{ \ketbra{\psi} : r_{\textsc{c}}(\ket{\psi})\leq k \},
	\end{equation}
	where $\operatorname{conv}$ stands for convex hull and $r_{\textsc{c}}$ is the coherence rank of $\ket{\psi}$, given by the number of non-zero coefficients in the basis-decomposition of $\ket{\psi}$. $\mathcal{C}_{1}$ is the set of fully incoherent states, given by density matrices that are diagonal in the computational basis, while $\mathcal{C}_{d} \equiv \mathscr{D(\mathscr H)}$ is the set of all states in the $d$-dimensional Hilbert space $\mathscr H$. It was shown in Ref.~\cite{Ringbauer2017} that these sets obey a strict hierarchy and that the amount of $k$-level coherence can be quantified by the robustness of multilevel coherence
	\begin{equation}
		R_{\mathcal{C}_k}(\rho) := \inf_{\tau\in \mathscr{D(H)}}\left\lbrace s \geq 0 : \frac{\rho + s \tau}{1+s} \in C_k \right\rbrace .
		\label{eq:RMC}
	\end{equation}
	Here $\tau$ is any density matrix. A state has coherence number $k$, if it can be decomposed into pure states which are superpositions of at most $k$ basis elements, while every decomposition must contain at least one such state. For $k=1$, this measure simply quantifies the total amount of coherence in the system and it can be computed efficiently for all $k$, given the density matrix. For $k=2$ and $k=3$ instead the results are shown in Fig.~\ref{fig:FigureS4}. Experimentally, we can quantify the multilevel coherence for a given density matrix using the following semi-definite program, see Ref.~\cite{Ringbauer2017} for details.
	
	\begin{equation}
		\label{Eq:RobustnessPrimal}
		\begin{array}{lll}
			R_{\mathcal{C}_k} (\rho) \,\,\, = \,\,\, & \min  \qquad & \Tr(\sum_{I\in\mathcal{P}_k}\tilde{\sigma}_{I}) -1 \\[6pt]
			& \textup{s.t.} & \tilde{\sigma}_{I} \geq 0 \quad\forall I\in \mathcal{P}_k\\[6pt]
			&  & P_{I}\tilde{\sigma}_{I}P_I = \tilde{\sigma}_I \quad\forall I\in \mathcal{P}_k\\[6pt]
			&  & \sum_{I\in\mathcal{P}_k} \tilde{\sigma}_{I} \geq \rho \, . \\
		\end{array}
	\end{equation}
	Here $\mathcal{P}_{k}$ is the set of all the $k$-element subsets of $\{1, 2, \ldots, d\}$, and $P_{I} := \sum_{i \in I} \ket{i}\bra{i}$

\end{document}